\documentclass[12pt]{article}

\usepackage{rotate}
\usepackage{graphicx}
\usepackage{amsmath}
\usepackage{amssymb}

\newcommand{\newc}{\newcommand}
\newc{\simgt}{\lower.7ex\hbox{$\;\stackrel{\textstyle>}{\sim}\;$}}
\newc{\simlt}{\lower.7ex\hbox{$\;\stackrel{\textstyle<}{\sim}\;$}}

\newenvironment{proposition}[1][Proposition]{\begin{trivlist}
\item[\hskip \labelsep {\bfseries #1}]}{\end{trivlist}}

\begin{document}

\begin{titlepage}
\begin{center}
{\hbox to\hsize{arXiv:0910.2730  \hfill   IZTECH-P2009-06}}

\bigskip
\vspace{3\baselineskip}

{\Large \bf  Vacuum Energy as the Origin of the Gravitational Constant}

\bigskip

\bigskip

{\bf  Durmu{\c s} A. Demir}\\
\smallskip

{ \small \it

Department of Physics, {\.I}zmir Institute of
Technology, TR35430, {\.I}zmir, Turkey}

\bigskip

{\tt  demir@physics.iztech.edu.tr}

\bigskip

\vspace*{.5cm}

{\bf Abstract}\\
\end{center}
\noindent
We develop a geometro-dynamical approach to the cosmological
constant problem (CCP) by invoking a geometry induced by the
energy-momentum tensor of vacuum, matter and radiation. The construction, which
utilizes the dual role of the metric tensor that it structures
both the spacetime manifold and energy-momentum tensor of the
vacuum, gives rise to a framework in which the vacuum energy
induced by matter and radiation, instead of gravitating, facilitates
the generation of the gravitational constant. The non-vacuum
sources comprising matter and radiation gravitate normally. At the
level of classical gravitation, the mechanism deadens the CCP yet quantum
gravitational effects, if can be strong in de Sitter space, can keep it existent.

\bigskip

\bigskip

\end{titlepage}

\section{Introduction and Motivation}
The matter-free spacetime is strictly flat, more precisely Ricci-flat,
unless its fabric is endowed with an intrinsic curvature $\Lambda_0$.
The ensuing spacetime curvature is governed by
\begin{eqnarray}
\label{field0}
R_{\alpha \beta}\left(\Gamma\right) - \frac{1}{2} g_{\alpha \beta} R\left(\Gamma\right) = - \Lambda_0\, g_{\alpha
\beta}
\end{eqnarray}
where $R_{\alpha \beta}$ is the Ricci tensor, $R$ the Ricci
scalar, and
\begin{eqnarray}
\label{levi-civita}
\Gamma^{\lambda}_{\alpha \beta} = \frac{1}{2} g^{\lambda \rho} \left( \partial_{\alpha} g_{\beta \rho}
+ \partial_{\beta} g_{\rho \alpha} - \partial_{\rho} g_{\alpha \beta}\right)
\end{eqnarray}
is the Levi-Civita connection. The constant curvature term
$\Lambda_0$, Einstein's cosmological constant \cite{einstein1},
represents an incalculable source of curvature, independent of
the entirety of matter and forces but gravity. As it stands, it
is a constant of Nature the value of which is to be determined
empirically.

In the presence of matter and radiation, the matter-free
gravitational field equations (\ref{field0}) change to
\begin{eqnarray}
\label{field1}
R_{\alpha \beta}\left(\Gamma\right) - \frac{1}{2} g_{\alpha \beta} R\left(\Gamma\right) = - \Lambda_0\, g_{\alpha
\beta} + 8 \pi G_N T_{\alpha \beta}
\end{eqnarray}
which is nothing but augmentation of (\ref{field0}) by the
energy-momentum tensor $T_{\alpha \beta}$ of matter and
radiation \cite{einstein2}. The prime idea in passing from
(\ref{field0}) to (\ref{field1}) is that gravitational
field is described by Poisson equation in the Newtonian limit.

In general, $T_{\alpha \beta}$ is computed from the quantum effective action for a given
background metric $g_{\alpha \beta}$ which necessarily
metamorphoses into a dynamical variable through (\ref{field1}).
On general grounds, energy-momentum tensor assumes the generic
form
\begin{eqnarray}
\label{en-mom}
T_{\alpha \beta} = - \texttt{E}\, g_{\alpha \beta} + {\texttt{t}}_{\alpha \beta}
\end{eqnarray}
where $\texttt{E}$ stands for the energy density of the vacuum
state, and  ${\texttt{t}}_{\alpha \beta}$ does for the
contributions of matter and radiation, collectively.
Replacement of this decomposition of $T_{\alpha \beta}$ into
(\ref{field1}) gives rise to an effective cosmological constant
\begin{eqnarray}
\label{lameff}
\Lambda_{\texttt{eff}} = \Lambda_0 + 8 \pi G_N \texttt{E}
\end{eqnarray}
which must nearly saturate the expansion rate of the Universe
\begin{eqnarray}
\label{cond}
\Lambda_{\texttt{eff}} \simlt H_0^2
\end{eqnarray}
since the non-vacuum mass contained in
$\texttt{t}_{\alpha \beta}$ is a small fraction of the critical
density. A number of independent observations \cite{astro1,astro2,astro3,astro4,astro5,astro6,astro7,astro8} have
determined $H_0$ to measure approximately $73.2\ {\rm
Mpc}^{-1}\, {\rm s}^{-1}\, {\rm km}$. This observational result
provides an experimental determination of
$\Lambda_{\texttt{eff}}$:
\begin{eqnarray}
\label{experi}
\Lambda_{\texttt{eff}}^{\texttt{exp}} \simeq 8 \pi G_N
\texttt{E}^{\texttt{exp}}
\end{eqnarray}
where $\texttt{E}^{\texttt{exp}} \simeq 3.25\times 10^{-47}\ {\rm GeV}^4$  \cite{astro1,astro6,astro7}
which, by just numerical coincidence, equals approximately $m_{\nu}^4$ with
$m_{\nu} \simeq 10^{-3}\ {\rm eV}$ being the neutrino mass.

If it were $\Lambda_0$ not $\Lambda_{\texttt{eff}}$, the bound
(\ref{cond}) would furnish, through the observational value of
$H_0$ quoted above, an empirical determination of $\Lambda_0$,
as for any other fundamental constant of Nature. The same does
not apply to $\Lambda_{\texttt{eff}}$, however. The reason is
that the vacuum energy density $\texttt{E}$, equaling the
zero-point energies of quantum fields plus enthalpy released by
various phase transitions, turns out to be characteristically
much larger than $\Lambda_{\texttt{eff}}^{\texttt{exp}}/ 8 \pi
G_N$. That this is the case can be illustrated by considering,
for instance, the electron which weighs next to neutrinos. The
electron loop gives sizeable contributions to $\texttt{E}$. The
smallest energy density it gives is electron mass per its cubic
Compton wavelength, and it is already much larger than
$\Lambda_{\texttt{eff}}^{\texttt{exp}}/ 8 \pi G_N$. Much
grosser than this is that the electron loop contributes to
$\texttt{E}$ by additional terms growing quadratically and
quartically with the ultraviolet cutoff. Consequently, the
known, experimentally confirmed fields and forces down to the
terascale, $M_{W} \sim {\rm TeV}$, are expected to induce a
vacuum energy density of order $M_{W}^4$  -- equaling the sum
total of energies deposited by parton-hadron and electroweak
phase transitions and by the quantal zero-point energies of
fields. This energy density, by itself, gives rise to a
centimeter-size Universe unless it is neutralized by the
$\Lambda_0$ contribution in (\ref{lameff}) -- a severe tuning
of numbers up to at least sixty decimal places. This immense
tuning worsens if the standard model of strong and electroweak
interactions extends beyond Fermi energies without a
suppression mechanism for $\texttt{E}$. One thus concludes
that, enforcement of $\Lambda_{\texttt{eff}}$ to obey
(\ref{cond}) gives rise to the biggest naturalness problem
plaguing both particle physics and cosmology -- the
cosmological constant problem. The CCP is a highly
inextricable perplexity a resolution of which is likely to be
found outside the framework set by the gravitational field
equations (\ref{field1}).

Over the decades, since its first solidification in \cite{ccp1,ccp2},
the CCP has been approached by putting forth various proposals
and interpretations, as listed and critically discussed in
\cite{nobbenhuis1,nobbenhuis2} (see also the review volumes \cite{review1,review2,review3,review4,review5}
and references therein). They each involve necessarily a
certain degree of speculative aspect in regard to going beyond
(\ref{field1}). These aspects involve postulating novel
symmetry arguments, relaxation mechanisms, modified
gravitational dynamics and statistical interpretations (see
\cite{1,2,3,4,5,6,7,8,9,10,11,12,13,14,15,16,17,18,19,20,21x,22,23,24,25,26,27,28,29,30,31,32,33,34,35,36,37,38,39,40,41,42,43,44,45,46,47,48,48p,49} for a partial list of recent work) as discussed
in \cite{nobbenhuis1,nobbenhuis2}. Excepting the nonlocal, acausal modification of
gravity implemented in
\cite{nima1,nima2} (see also \cite{33}) and the anthropic approach \cite{anthro}, most of
the solutions proposed for the CCP seem to overlook the already
existing vacuum energy density ${\mathcal{O}}\left({\rm
TeV}^{4}\right)$ induced by the known physics down to the
terascale \cite{weinberg1,weinberg2,weinberg3}. Indeed, any resolution of the CCP,
irrespective of how speculative it might be, must, in the first
place, provide an understanding of how this existing
energy component to be tamed.

Having stated the problem, we are at the stage to lay out the
germ of the mechanism to be proposed in the present work. The
alleged mechanism rests on finding a sensible answer to the
question: {\it Can one excogitate a way, different than in
(\ref{field1}), of incorporating $T_{\mu \nu}$ into the
matter-free gravitational field equations (\ref{field0}) while
keeping all the successes of the Einstein field equations for
non-vacuum sources $\texttt{t}_{\alpha \beta}$ yet naturalizing
the effects of the vacuum energy $\texttt{E}$ ?} The answer is
affirmative. The method, as will be described in Sec. 2 below,
involves incorporation of matter and radiation
into (\ref{field0}) by replacing the metric $g_{\alpha \beta}$
by a general tensor field $\mathbb{T}_{\alpha \beta}$ -- to be
related to the energy-momentum tensor $T_{\alpha \beta}$.
In Sec. 3 it will be shown that the dynamics in Sec. 2
follow from an action principle. Sec. 4 is devoted to a
critical discussion of the method  analyzed in Sec. 2.
In Sec. 5 we conclude.

\section{An Alternative Route}
For incorporating matter and radiation into (\ref{field0})
in a way desirably free of the CCP, we propose an alternative approach
wherein the vacuum energy density $\texttt{E}$ is offloaded
from the effective cosmological constant  $\Lambda_{\texttt{eff}}$
in (\ref{lameff}). We lay out the
proposal by moving from abstract to concrete, where concreteness
will be judged on the basis of the physical relevance of the resulting
dynamical equations in regard to (\ref{field1}). The `propositions'
below should not be taken in strict mathematical sense; they are merely a
logically complete set of physical statements which will
form the mechanism proposed.

$\S$ The primary statement of the proposal is as follows.
\begin{proposition} {\bf 1.}
{\it Let
\begin{eqnarray}
\label{replace}
\mathbb{T}_{\alpha \beta} = \Lambda g_{\alpha \beta} + \Theta_{\alpha \beta}
\label{propose1}
\end{eqnarray}
be a generic tensor field with $\Lambda$ being a nonzero constant and $\Theta_{\alpha \beta}$
a symmetric tensor field with well-defined matrix inverse. Then replacement of the metric tensor $g_{\alpha\beta}$
in the matter-free gravitational field equation (\ref{field0}) by $\mathbb{T}_{\alpha\beta}$
gives rise to a novel field equation possessing the fundamental property that
the $\Lambda$ term in $\mathbb{T}_{\alpha \beta}$ does give no contribution, additive or
otherwise, to the original curvature source $\Lambda_0$.}
\end{proposition}
The proof starts with the field equations
\begin{eqnarray}
\label{eqn2}
\mathbb{R}_{\alpha \beta}\left(\between\right) - \frac{1}{2}
\mathbb{T}_{\alpha \beta} \left(\mathbb{T}^{-1}\right)^{\mu
\nu} \mathbb{R}_{\mu \nu}\left(\between\right) = - \frac{\Lambda_0}{\Lambda} \mathbb{T}_{\alpha \beta}
\end{eqnarray}
arising from (\ref{field0}) after replacing $g_{\alpha \beta}$ therein by (\ref{replace}). It is worthy of
noting that this equation uniquely reduces to (\ref{field0}) as $\Theta_{\alpha\beta}\rightarrow 0$.
In fact, the factor of $1/\Lambda$ at the right-hand side arises for this very reason.

As a direct consequence of (\ref{replace}), the connection changes from the Levi-Civita
connection (\ref{levi-civita}) to
\begin{eqnarray}
\label{eqn3}
\between^{\lambda}_{\alpha \beta} &=& \frac{1}{2} \left(\mathbb{T}^{-1}\right)^{\lambda
\nu} \left( \partial_{\alpha} \mathbb{T}_{\beta\nu} + \partial_{\beta} \mathbb{T}_{\nu \alpha} - \partial_{\nu} \mathbb{T}_{\alpha \beta}\right)
= \Gamma^{\lambda}_{\alpha \beta} + \Delta^{\lambda}_{\alpha \beta}
\end{eqnarray}
where
\begin{eqnarray}
\label{tensor-conn}
\Delta^{\lambda}_{\alpha \beta} = \frac{1}{2} \left(\mathbb{T}^{-1}\right)^{\lambda
\nu} \left( \nabla_{\alpha} \mathbb{T}_{\beta\nu} + \nabla_{\beta} \mathbb{T}_{\nu \alpha} - \nabla_{\nu} \mathbb{T}_{\alpha \beta}\right)
\end{eqnarray}
is a tensor field involving covariant derivatives with respect
to $\Gamma^{\lambda}_{\alpha \beta}$. This tensorial connection
identically vanishes when $\Theta_{\alpha \beta} = 0$.
Therefore, it is a sensitive probe of the
covariantly-nonconstant part $\Theta_{\alpha \beta}$ of
$\mathbb{T}_{\alpha \beta}$. As suggested by (\ref{tensor-conn}),
$\Delta^{\lambda}_{\alpha \beta}$ exhibits a rather
specific dependence on $\Theta_{\alpha \beta}$ and $\Lambda$:
\begin{eqnarray}
\label{tensor-conn-depend}
\Delta^{\lambda}_{\alpha \beta}\left(\Theta, \Lambda\right) = \Delta^{\lambda}_{\alpha \beta}\left(\frac{\Theta}{\Lambda}\right)
\end{eqnarray}
from which it follows that deviations from the decomposition of
$\mathbb{T}_{\alpha \beta}$ defined in (\ref{replace}),
\begin{eqnarray}
\label{additive}
\Theta_{\alpha \beta} \rightarrow \delta\Lambda g_{\alpha
\beta} + \Theta^{\prime}_{\alpha \beta}\,,
\end{eqnarray}
with $\delta\Lambda$ being a  constant increment in $\Lambda$, is reflected as
\begin{eqnarray}
\label{tensor-conn-nonadditive}
\Delta^{\lambda}_{\alpha \beta}\left(\frac{\Theta}{\Lambda}\right) \rightarrow
\Delta^{\lambda}_{\alpha \beta}\left(\frac{\Theta^{\prime}}{\Lambda + \delta\Lambda}\right)
\end{eqnarray}
which bears no structural change compared to
(\ref{tensor-conn-depend}).

The geometry induced by $\mathbb{T}_{\alpha\beta}$ is manifestly
non-Riemannian, as explicated by the split structure of
$\between^{\lambda}_{\alpha \beta}$ in (\ref{eqn3}). In fact, in response
to this structure,  the curvature tensor in (\ref{eqn2}) parts into two
\begin{eqnarray}
\label{split} \mathbb{R}_{\alpha \beta}\left(
\between\right) = R_{\alpha \beta}\left(\Gamma\right) +
{\cal{R}}_{\alpha \beta}\left(\Delta\right)
\end{eqnarray}
which differs from $R_{\alpha \beta}\left(\Gamma\right)$
operating in (\ref{field0}) by
\begin{eqnarray}
\label{riemannextra} {\cal{R}}_{\alpha \beta}\left(\Delta\right) = \nabla_{\mu}
\Delta^{\mu}_{\alpha \beta} - \nabla_{\beta} \Delta^{\mu}_{\mu
\alpha} + \Delta^{\mu}_{\mu
\nu} \Delta^{\nu}_{\alpha \beta} - \Delta^{\mu}_{\beta
\nu} \Delta^{\nu}_{\alpha \mu}
\end{eqnarray}
which is a symmetric tensor field induced by $\Delta$ alone.
This is a quasi-curvature tensor in that it is not generated by
commutators of $\nabla$ or $\nabla^{\between}$, and hence, it
does not obey the Bianchi identities. The functional dependence
of $\Delta$ on $\Theta$, given in (\ref{tensor-conn-depend}),
guarantees that ${\cal{R}}_{\alpha \beta}\left(\Delta\right)$
depends solely on $\Theta_{\alpha \beta}/\Lambda$ and its
derivatives. More important, however, is the fact that
${\cal{R}}_{\alpha \beta}\left(\Delta\right)$, by any means,
does neither possess nor develop any covariantly-constant part
due to the scaling property of $\Delta$ in (\ref{tensor-conn-nonadditive}).

Consequently, given the structure of ${\cal{R}}_{\alpha \beta}\left(\Delta\right)$,
and given also the scaling property (\ref{tensor-conn-nonadditive})
of the tensorial connection $\Delta^{\lambda}_{\alpha\beta}$, one arrives at the
firm conclusion that the field equations (\ref{eqn2}) possess {\it one and  only one}
covariantly-constant source which is $\Lambda_0$. Therefore, the matter-free field
equation (\ref{field0}) and the proposed one (\ref{eqn2}) do have the same
cosmological term. In other words, presence of matter and radiation, irrespective of
how large a vacuum energy density is deposited, does not change the cosmological
constant $\Lambda_0$.

This completes the proof of {\bf Proposition 1}. $\blacksquare$

$\S\S$ The field equations (\ref{eqn2}) is an abstract one in that it bears no
indicant of any connection to the gravitational field dynamics in the presence of matter and radiation.
It is just a dynamical equation for $\mathbb{T}_{\alpha\beta}$. For it to
gain a concrete overtone, one must, in the first place, determine the requisite
relation between $\mathbb{T}_{\alpha\beta}$ and $T_{\alpha\beta}$. To this end,
it proves convenient to rearrange (\ref{eqn2}) by using (\ref{split}) to get
\begin{eqnarray}
\label{eom}
R_{\alpha \beta}\left(\Gamma\right) - \frac{1}{2} g_{\alpha \beta} R\left(\Gamma\right) &=& -\frac{\Lambda_0}{\Lambda} \mathbb{T}_{\alpha \beta}
-\frac{1}{2} \left[ g_{\alpha \beta} g^{\mu \nu} - \mathbb{T}_{\alpha \beta} \left(\mathbb{T}^{-1}\right)^{\mu
\nu} \right]{R}_{\mu \nu}\left(\Gamma\right)\nonumber\\
&-&\left[{\cal{R}}_{\alpha \beta}\left(\Delta\right) - \frac{1}{2} \mathbb{T}_{\alpha \beta} \left(\mathbb{T}^{-1}\right)^{\mu
\nu} {\cal{R}}_{\mu \nu}\left(\Delta\right)\right]
\end{eqnarray}
whose right-hand side, upon an appropriate relation between $\mathbb{T}_{\alpha\beta}$ and $T_{\alpha\beta}$,
must reduce to that of (\ref{field1}), excluding the vacuum contribution, at least as the leading structure.
The dynamical equations resulting from (\ref{eom}) will be judged on the basis of physical consistency and phenomenological relevance.

Before ascertaining the appropriate relation between $\mathbb{T}_{\alpha\beta}$ and $T_{\alpha\beta}$,
one notes a fundamental aspect of (\ref{eom}):

\begin{proposition} {\bf 2.}
{\it The matter-nonfree gravitational field equations which will spring from (\ref{eom}) will have
$\Lambda_0$ as the only source for cosmological constant. Therefore,
$\Lambda_0$ remains isolated and is empirically determined from cosmological
observations, and this determination involves no fine-tuning of distinct
energy sources.}
\label{propose1p}
\end{proposition}
The proof relies on {\bf Proposition 1} itself. The
gravitational field equations, whether they are admissible or not,
will contain $\Lambda_0$ as the only covariantly-constant curvature
source, and matter sector will give no contributions to it thanks
to (\ref{tensor-conn-nonadditive}) as well as the $\Lambda_0/\Lambda$
factor at the right-hand side of (\ref{eom}). As $\Lambda_0$
receives no contribution from matter sector, it is by itself the
cosmological term, and it must saturate the observational result
\begin{eqnarray}
\label{lam0}
\Lambda_0 \simeq 8 \pi G_N \texttt{E}^{\texttt{exp}}
\end{eqnarray}
where $\texttt{E}^{\texttt{exp}} \simeq m_{\nu}^{4}$, as indicated by the astrophysical
observations \cite{astro1,astro2,astro3,astro4,astro5,astro6,astro7,astro8}.
The curvature source $\Lambda_0$ which continues to be the only cosmological
term  with (as in (\ref{eom})) or without (as in (\ref{field0})) matter and
radiation, its empirical determination from observations, no different than
fixing Newton's constant or gauge couplings or any other model parameter by
using the experimental data, involves no tuning of parameter values.

This completes the proof of {\bf Proposition 2}. $\blacksquare$

If the field equation (\ref{eom}) for $\mathbb{T}_{\alpha \beta}$ is to
have anything to do with the gravitational field dynamics in the presence
of matter and radiation,  $\mathbb{T}_{\alpha\beta}$ and $T_{\alpha\beta}$
must be put in relation in an appropriate way. There is no {\it a priori}
telling of what this alleged relation should be. It could be a local as well
as a nonlocal relation. It could be a linear as well as a nonlinear relation.
There are all sorts of variations one can consider. In this work, consideration will be
on two classes of relations between $\mathbb{T}_{\alpha\beta}$ and $T_{\alpha \beta}$:
{\it local} and {\it nonlocal} relations. As will be proven below, these two will
reveal fundamental aspects of the desired relation.

\begin{proposition} {\bf 3.}
{\it It is impossible to achieve a resolution for the CCP through a local, linear
relation if $\Lambda_0$ is to saturate the experimental result (\ref{experi}),
and if the gravitational constant is to be generated correctly. The nonlinear,
local relations, on the other hand, fail to yield correct gravitational dynamics
due to the presence of higher powers and derivatives of $T_{\alpha\beta}$ in
the resulting equations.}
\label{propose2}
\end{proposition}
The proof starts by taking
\begin{eqnarray}
\label{linear-loc}
\mathbb{T}_{\alpha\beta} =C_l T_{\alpha \beta}
\end{eqnarray}
with which the right-hand side of (\ref{eom}) becomes a function of $T_{\alpha \beta}$ alone.
$C_l$ is a constant. For recovering correct gravitational dynamics in the sense of (\ref{field1}), one imposes
$\Lambda_0 \simeq \Lambda^{\texttt{exp}}_{\texttt{eff}}$ and, by using $-\left(\Lambda_0/\Lambda\right)\Theta_{\alpha\beta} = \left(\Lambda_0/\texttt{E}\right)\texttt{t}_{\alpha\beta}$, equates the coefficient of $\texttt{t}_{\alpha \beta}$, $\Lambda_0/\texttt{E}$, to $8 \pi G_N$.
This requires $\texttt{E} \simeq \texttt{E}^{\texttt{exp}}\simeq m_{\nu}^4$, independent of $C_l$.
Startlingly, this result is nothing but the CCP itself \cite{ccp1,ccp2,weinberg1,weinberg2}.

For breaking this impasse, one can try a more general structure $\left[f(T)\right]_{\alpha \beta}$
instead of the linear one (\ref{linear-loc}). Expanding this tensor structure in powers of $T_{\alpha \beta}$
around the origin and identifying the coefficient of $\texttt{t}_{\alpha \beta}$ with the gravitational constant,
one finds
\begin{eqnarray}
\label{nonlin-loc}
\mathbb{T}_{\alpha\beta}=\left[f(T)\right]_{\alpha \beta} = C_n \left[\exp\left\{ - \frac{T}{\texttt{E}^{\texttt{exp}}}\right\}\right]_{\alpha \beta}
\end{eqnarray}
where $C_n$ is a constant. It is obvious that unless, in size, $T_{\alpha \beta}$ is small compared to
$\texttt{E}^{\texttt{exp}}\simeq m_{\nu}^4$ this function does not admit a power series expansion. In fact, what it
clearly shows is that, unless the CCP is solved, construction of an admissible local relation
between $\mathbb{T}_{\alpha\beta}$ and $T_{\alpha\beta}$ is difficult if not impossible. The situation gets
even worse if one enforces the right-hand side of (\ref{eom}) to have vanishing divergence as its
left-hand side is divergence-free by the Bianchi identities.  Indeed, inclusion of this condition would greatly reduce the
admissible relations between $\mathbb{T}_{\alpha\beta}$ and $T_{\alpha\beta}$. To see how serious
this condition is, one notes that both (\ref{linear-loc}) and (\ref{nonlin-loc}) have vanishing divergence;
however, neither of them can nullify the divergence of the right-hand side of (\ref{eom}).

This completes the proof of {\bf Proposition 3}. ${\blacksquare}$

\begin{proposition} {\bf 4.}
{\it A linear, causal, nonlocal relation between $\mathbb{T}_{\alpha\beta}$ and $T_{\alpha\beta}$, in the limit
of large $\Lambda$, gives rise to correct gravitational dynamics up to a nonlocal
${\cal{O}}\left(1/\Lambda^2\right)$ remainder. This remainder is not divergence-free, and hence,
causes inconsistency in regard to Bianchi identities. This inconsistency can be cured by a nonlocal,
nonlinear relation which might be constructed by perturbing the linear relation order by order in $1/\Lambda$.}
\label{propose3}
\end{proposition}
The proof follows the view point that the linear-in-$\texttt{t}_{\alpha\beta}$ term, necessary for recovering (\ref{field1}),
can actually come from the third term in (\ref{eom}),  which involves ${\cal{R}}_{\alpha \beta}\left(\Delta\right)$,
and hence, derivatives of $\Theta_{\alpha \beta}$. The cosmological constant is still fed by the
first term in (\ref{eom}); however, its $\Theta_{\alpha\beta}$ part becomes subdominant due
to $1/\Lambda$ suppression in front. In accord with  {\bf Proposition 4}, when $\Lambda$
is much larger than $\Theta_{\alpha\beta}$ in size, the third term in (\ref{eom}) takes the form
\begin{eqnarray}
\label{expand-rhs}
{\cal{R}}_{\alpha \beta}\left(\Delta\right) - \frac{1}{2} \mathbb{T}_{\alpha \beta} \left(\mathbb{T}^{-1}\right)^{\mu
\nu} {\cal{R}}_{\mu \nu}\left(\Delta\right) &=&\frac{1}{2 \Lambda} \left(G^{-1}\right)_{\alpha\beta\mu\nu} \mathbb{T}^{\mu \nu}\nonumber\\ &+& {\cal{O}}\left(\frac{\nabla\Theta \nabla\Theta}{\Lambda^2}, \frac{\Theta\nabla\nabla\Theta}{\Lambda^2}\right)
\end{eqnarray}
where, at the right-hand side, the operator acting on $\mathbb{T}^{\mu\nu}$ reads as
\begin{eqnarray}
\left(G^{-1}\left(\nabla\right)\right)_{\alpha\beta\mu\nu} &=& \nabla_{\mu}\nabla_{\alpha} g_{\nu\beta}
+ \nabla_{\mu}\nabla_{\beta} g_{\nu\alpha} - \nabla_{\mu}\nabla_{\nu} g_{\alpha\beta}\nonumber\\
&-& \nabla_{\alpha} \nabla_{\beta} g_{\mu \nu} - \Box g_{\mu \alpha} g_{\nu \beta}  +
\Box g_{\mu \nu} g_{\alpha \beta}
\end{eqnarray}
up to the additive terms needed for symmetrization with respect to $\left(\alpha,\beta\right)$ and $\left(\mu,\nu\right)$.
This operator is clearly the inverse of the massless spin-2 field propagator
$G_{\alpha \beta \mu \nu}\left(\nabla\right)$ in the background metric $g_{\alpha \beta}$
\cite{neuwein1,neuwein2}. For recovering the Einstein field equations (\ref{field1}),
one must require (\ref{expand-rhs}) to be proportional to the energy-momentum tensor
of matter and radiation $\texttt{t}_{\alpha \beta}$ (as it cannot involve any
covariantly-constant piece involving $\texttt{E}\ g_{\alpha \beta}$), and we take it, on
the basis of linearity, to be  equal to $- \texttt{t}_{\alpha \beta}/2 \Lambda$, that is,
\begin{eqnarray}
\label{pertur-eq}
\left(G^{-1}\left(\nabla\right)\right)_{\alpha\beta\mu\nu} \mathbb{T}^{\mu \nu} = - \texttt{t}_{\alpha \beta}
\end{eqnarray}
where the minus sign is necessitated by (\ref{eom}). Integration of this relation
necessarily involves a constant of integration in the sense of covariant derivatives. This covariantly-constant tensor
must be proportional to the vacuum energy-momentum tensor $-\texttt{E} g_{\alpha \beta}$.
Consequently, the requisite relation between $\mathbb{T}_{\alpha \beta}$ and  $T_{\alpha \beta}$, within
the linearized regime in (\ref{expand-rhs}), turns out to be
\begin{eqnarray}
\label{relate}
\mathbb{T}_{\alpha \beta} = - \texttt{L}^2 \texttt{E} g_{\alpha \beta} - G_{\alpha \beta \mu \nu}\left(\nabla\right) \texttt{t}^{\mu \nu}
\end{eqnarray}
from which it follows, in view of (\ref{replace}), that
\begin{eqnarray}
\Lambda \equiv - \texttt{L}^2 \texttt{E}\;,\;\;
\Theta_{\alpha \beta} \equiv - G_{\alpha \beta \mu \nu}\left(\nabla\right) \texttt{t}^{\mu \nu}
\end{eqnarray}
where $\texttt{L}^2$, an algebraic area scale, necessarily
arises for dimensionality reasons. From (\ref{relate}) it directly follows that
$\mathbb{T}_{\alpha \beta}$ is related to $T_{\alpha \beta}$
{\it linearly} by the construction,
{\it nonlocally} by the fact that  $\mathbb{T}_{\alpha \beta}$ at a given point involves $\texttt{t}_{\mu\nu}$
in entire spacetime as propagated by $G_{\alpha \beta \mu \nu}$, and
{\it causally} by the fact that $G_{\alpha \beta \mu \nu}$ is causal in
chronologically structured spacetimes. The nonlocality here reminds one at once the one arising in \cite{nima1,nima2};
however, one notices that dynamical structure here is entirely different in regard to (\ref{eom}).

Having the relation (\ref{relate}) at hand, the equation of motion (\ref{eom}) takes the form
\begin{eqnarray}
\label{result}
R_{\alpha \beta}\left(\Gamma\right) - \frac{1}{2} g_{\alpha \beta} R\left(\Gamma\right) &=& \frac{1}{2 \Lambda} {\texttt{t}}_{\alpha
\beta} - {\Lambda_0} g_{\alpha \beta} + \frac{\Lambda_0}{\Lambda}
\left(\Theta_{\alpha \beta} - \frac{1}{2} g^{\mu \nu} \Theta_{\mu \nu} g_{\alpha \beta}\right)\nonumber\\
&+&{\cal{O}}\left(\frac{\Lambda_0}{\Lambda^2} \Theta^2, \frac{1}{\Lambda^2}\Theta^3, \frac{1}{\Lambda^2} \Theta \texttt{t} \right)
\end{eqnarray}
whose first term, proportional to $\texttt{t}_{\alpha \beta}$,
manifestly imposes the identification
\begin{eqnarray}
\label{newton}
8 \pi G_{N} \equiv \frac{1}{\overline{M}_{Pl}^2} = \frac{1}{2 \Lambda}
\end{eqnarray}
so that the entire material existence, excluding the vacuum,
gravitates, as in General Relativity, via the Einstein field equations (\ref{field1}).
This definition of the Newton's constant, through
(\ref{relate}), inherently presumes that $\texttt{L}^2$ and
$\texttt{E}$ are of the same sign. Moreover, it is obvious that
the power series expansion in $\Theta/\Lambda$, employed to
obtain (\ref{expand-rhs}), and in turn, to get (\ref{result}) from (\ref{eom}),
is fully justified.

The gravitational field equations (\ref{result}) possess certain features deserving a separate, detailed discussion.
\begin{enumerate}
\item
The area scale $\texttt{L}^2$ is determined by the
vacuum energy density through (\ref{newton}). It possesses the physical extrema
\begin{eqnarray}
\label{lkare}
&&\texttt{E} \simeq M_{W}^4 \Longrightarrow \texttt{L}^2 \simeq \frac{1}{m_{\nu}^{2}}\nonumber\\
&&\texttt{E} \simeq \overline{M}_{Pl}^4 \Longrightarrow \texttt{L}^2 \simeq \frac{1}{\overline{M}_{Pl}^{2}}
\end{eqnarray}
which directly follow from (\ref{newton}). These limiting values show that the length scale $\texttt{L}$ ranges
from the neutrino Compton wavelength down to the Planck length. The vacuum energy density cannot
fall below $M_W^4$ significantly, if electroweak breaking is to generate
the observed gauge boson masses. Its value gets stabilized at or above $M_W^4$ if contributions of higher-frequencies
are nullified by some mechanism such as the global supersymmetry \cite{nobbenhuis1,nobbenhuis2}. In case Nature
does not utilize such a mechanism, vacuum energy density rises to $\overline{M}_{Pl}^4$ and, correspondingly,
$\texttt{L}$ falls down to the Planck length. This particular value of vacuum energy density gives rise to a theory
with a single parameter, $\overline{M}_{Pl}$. This elegant setup, tantalizingly, corresponds to the
worst case in General Relativity: In the framework of (\ref{field1}), with $\texttt{E} \simeq \overline{M}_{Pl}^4$,
the fine-tuning needed to satisfy (\ref{cond}) rises to 120 decimal places.

\item

The nonlocal nature of (\ref{relate}) spreads all over the right-hand side of (\ref{eom}).
Though the leading linear-in-$\texttt{t}_{\alpha\beta}$ term in (\ref{result}) is inherently
local, the subleading ones are not. The violation of locality, under (\ref{newton}) and (\ref{lam0}),
turns out to be exceedingly small, however. The reason is that the third term in
(\ref{result}), which is clearly nonlocal, is dressed by extreme
${\cal{O}}\left(m_{\nu}^4/\overline{M}_{Pl}^4\right)$ suppression. The remaining
${\cal{O}}\left(1/\Lambda^2\right)$ terms, which involve derivatives of $\Theta_{\alpha\beta}$,
also give nonlocal contributions. These sources of nonlocality are exceedingly small,
and thus, the world as we see behaves local to a good approximation.

\item
The divergence of the left-hand side of (\ref{field1}) vanishes by
the differential Bianchi identity, and that of its right-hand side vanishes by
energy-momentum conservation. This feature is geometrical for the
left-hand side and kinematical for the right-hand side. This is
the situation in General Relativity. In the present approach, however,
the problem gains a dynamical nature in regard to (\ref{eom});
the divergence of its right-hand side involves gradients of
$\mathbb{T}_{\alpha \beta}$, $R_{\alpha\beta}\left(\Gamma\right)$ and ${\cal{R}}_{\alpha\beta}\left(\Delta\right)$
with no obvious telling of how it can identically vanish. Given this variety in the structures involved,
enforcing the divergence to vanish gives rise to non-trivial constraints on
$\mathbb{T}_{\alpha\beta}$. $\mathbb{T}_{\alpha\beta}$ must be related to $T_{\alpha \beta}$
in such a way that the right-hand side of (\ref{eom}) becomes divergence-free. Obviously, one
can find a wiser solution which already satisfies this constraint at the exact, nonlinear level.
Leaving aside this possibility, all one can do is to iterate expansion in $\Theta/\Lambda$ to
higher powers and impose vanishing of divergence order by order in perturbation theory. That
this procedure can work is already revealed by (\ref{result}) wherein the lowest order terms,
the first two terms at the right-hand side, possess vanishing divergence. For making
divergence to vanish at the next order in $\Theta/\Lambda$, one iterates the linear relation
between $\mathbb{T}_{\alpha\beta}$ and $T_{\alpha \beta}$ in (\ref{relate}) to include ${\cal{O}}\left(1/\Lambda\right)$ terms
\begin{eqnarray}
\mathbb{T}_{\alpha\beta} = \mathbb{T}_{\alpha\beta}\left(\mbox{in\, Eq.}\; \ref{relate}\right) + \frac{1}{\Lambda} \texttt{Q}_{\alpha\beta}
\end{eqnarray}
where $\texttt{Q}_{\alpha\beta}$ is an appropriate tensor structure which makes ${\cal{O}}\left(1/\Lambda^2\right)$ terms divergence-free.
Clearly, $\texttt{Q}_{\alpha\beta}$ develops both local and nonlocal dependencies on $T_{\alpha\beta}$. This iteration
continues to higher powers of $T_{\alpha\beta}$ by including local as well as nonlocal contributions according to the terms arising
in the expansion (\ref{expand-rhs}). This iterative procedure, as has also been discussed in \cite{nima1} for a similar problem, can make the right-hand
side of (\ref{eom}) divergence-free order by order in perturbation theory. The final result is that $\mathbb{T}_{\alpha\beta}$ develops
a nonlocal and nonlinear relation with $T_{\alpha\beta}$, and the right-hand side of (\ref{result}) is still nonlocal yet divergence-free.
\end{enumerate}
The analysis above completes the proof of {\bf Proposition 4}. $\blacksquare$

$\S\S\S$
The equations of motion (\ref{eqn2}) have been obtained from
the matter-free gravitational field equations (\ref{field0})
by simply replacing the metric tensor$g_{\alpha \beta}$ by
$\mathbb{T}_{\alpha \beta}$. This pragmatic approach
is eventually justified {\it a posteriori} via the propositions stated
and proven above. Nevertheless, it is desirable to derive
(\ref{eqn2}) from an invariant action, as discussed below.

\begin{proposition} {\bf 5.}
{\it The equation of motion (\ref{eqn2}) for $\mathbb{T}_{\alpha\beta}$ follows
from an action principle.}
\label{propose4}
\end{proposition}
The proof proceeds by explicit construction. The requisite action,
which does not need to be unique, may be modeled as
\begin{eqnarray}
\label{basic-action}
I&=& N \int d^4 x\ \sqrt{\widetilde{\mathbb{T}}} \Bigg[ \Lambda \left(\widetilde{\mathbb{T}}^{-1}\right)^{\alpha
    \beta} \mathbb{R}_{\alpha \beta}\left(\widetilde{\between}\right) - 2 \Lambda_0\nonumber\\ &+& \lambda_{g m} \left( \widetilde{\mathbb{T}}^{-1} -
    {\mathbb{T}}^{-1}\right)^{\alpha \beta} \left(\widetilde{\mathbb{T}} - {\mathbb{T}}\right)_{\alpha \beta}
\Bigg] + I_{matter}\left[g, \psi\right]
\end{eqnarray}
where $N$ is a  normalization constant, $\lambda_{g m}$ is a Lagrange multiplier,
$\mathbb{T}_{\alpha \beta}$ is as defined in (\ref{replace}), $\psi$ collectively denotes the
matter fields, and $I_{matter}\left[g, \psi\right]$ is the quantum effective action
for the matter sector. The matter action, constructed on the background metric $g_{\alpha\beta}$,
involves quantum corrections from matter loops as well as non-renormalizable interactions
from physics at short distances. It should be emphasized that the matter action, at the
classical level, possesses no single term without the matter fields, that is, {\it it vanishes identically for $\psi = 0$}.
This restriction does, of course, not modify interactions of fields with the background geometry:
They can interact via their kinetic terms and other means such as non-minimal coupling of scalar fields
to curvature scalar $R\left(\Gamma\right)$ of the background geometry. These interactions are
as in the General Relativity.

The gravitational sector in (\ref{basic-action}) can be augmented by the curvature invariants
$\left(\mathbb{T}^{-1}\right)^{\alpha \beta} \mathbb{R}_{\alpha \beta}$, $\left(\mathbb{T}^{-1}\right)^{\alpha \mu}
\left(\mathbb{T}^{-1}\right)^{\beta \nu} \mathbb{R}_{\alpha \beta} \mathbb{R}_{\mu \nu}$, and the like.
These higher-derivative contributions are to be suppressed by some large mass scale, and in fact, they must be absent
if the ghosts are to be eliminated from the spectrum.

The action above is constructed within the Palatini formulation
\cite{palatini1,palatini2} in that connection $\widetilde{\between}^{\lambda}_{\alpha
\beta}$ and metric $\widetilde{\mathbb{T}}_{\alpha \beta}$ are completely
independent geometrical quantities to start with. In fact, as
it stands, it is a generalization of the Born-Infeld action \cite{infeld}.
Extremization of the action with respect to $\lambda_{g m}$ gives
\begin{eqnarray}
\widetilde{\mathbb{T}}_{\alpha \beta} = {\mathbb{T}}_{\alpha \beta}
\end{eqnarray}
in accord with (\ref{replace}). On the other hand, its extremization
with respect to $\left({\mathbb{T}}^{-1}\right)^{\alpha \beta}$
returns precisely the equations of motion (\ref{eqn2}), and the one
with respect to $\widetilde{\between}^{\lambda}_{\alpha \beta}$ gives
\begin{eqnarray}
\label{eqn1}
\nabla^{\widetilde{\between}}_{\lambda} \left(\sqrt{\mathbb{T}} \left(\mathbb{T}^{-1}\right)^{\alpha \beta}\right) = 0
\end{eqnarray}
which is directly solved by the connection coefficients in (\ref{eqn3}).
Finally, extremization of the whole action with respect to $g_{\alpha \beta}$ returns
the energy-momentum tensor $T_{\alpha \beta}$, and the one
with respect to matter fields $\psi$ gives the usual field equations,
as they must.

One notices that the action functional above does not encode the relation between $\mathbb{T}_{\alpha\beta}$
and $T_{\alpha\beta}$. The functional method above presumes that the alleged relation must be supplied
externally in order to get an admissible gravitational dynamics, as in {\bf Proposition 4}.

This completes the proof of {\bf Proposition 5}. $\blacksquare$

$\S\S\S\S$ Throughout the  {\bf Propositions 1}-{\bf 5}, gravity
has been assumed to be an inherently classical phenomenon whose primary role being establishment
of the background geometry for the quantal matter. However, for integrity and completeness,
it is necessary to determine if the whole mechanism, in particular, the setup in (\ref{result})
is stable under quantum gravitational effects.

\begin{proposition} {\bf 6.}
{\it The CCP-solving setup in (\ref{result}) receives only nonviolent quantum gravitational
corrections, and the vacuum energy density $\texttt{E}$ picks out a value around $\overline{M}_{Pl}^4$
unless one disproves the conjectures that the de Sitter gravity is essentially classical and
de Sitter space does not support supersymmetry.}
\label{propose5}
\end{proposition}
The proof starts by noting that $\Lambda_0 > 0$ by the astrophysical observations \cite{astro1,astro6,astro7},
and thus, the background geometry, described by (\ref{field0}) which coincides with (\ref{result})
in the matter-free regime,  is de Sitter space. The de Sitter space, having its asymptotia in the
past and future with no notion of spatial infinity, possesses the curvature radius $\ell = \left(3/\Lambda_0\right)^{1/2}$.
The area of the horizon seen by causal observers is ${\cal{A}}= 4 \pi \ell^2 = 12 \pi/\Lambda_0$. Consequently,
the entropy of the empty de Sitter space is ${\cal{S}}={\cal{A}}/4G_N= 3\pi/\left(G_N \Lambda_0\right)$, and
it is conjectured that this entropy continues to be the maximal one when matter and radiation are present \cite{bousso1}.
In accord with statistical mechanical interpretation, this entropy must equal to the logarithm of the number of
quantum microstates of the de Sitter space, and hence, the cosmological constant $\Lambda_0$, which receives no
contribution from the matter sector according to previous propositions, is nothing but a measure of the dimension of
the Hilbert space of the quantum de Sitter gravity \cite{banks1,witten1}. For nonvanishing $\Lambda_0$, as enforced
by astrophysical observations \cite{astro1,astro6,astro7}, the dimension of the Hilbert space is finite, and this implies that
de Sitter gravity cannot be quantized \cite{witten1,onemli1}; it must result
from a more fundamental theory that predicts $\Lambda_0$. However, there are arguments that classical de Sitter space
cannot arise from compactifications of string or M theories \cite{witten1,onemli2}.  In other words, for the experimental value of $\Lambda_0$
in (\ref{experi}), $G_N \Lambda_0\simeq 10^{-120}$ is sufficiently small to admit a perturbative string theory or supergravity
description yet there is no known example of compactification of these theories into de Sitter space. Besides,
having no classical compactification it is not clear how it can exist at the quantum level \cite{witten1}. Nevertheless,
there are indications that semiclassical de Sitter space (suppressed quantum fluctuations) can originate from quantum excitations at the
Planck scale, nonperturbatively \cite{onemli3,onemli4}. These arguments, at least in perturbative regime, guarantee that the de Sitter
gravity is essentially classical, and thus, there arise no tensor fluctuations to induce a quantum gravitational vacuum energy.
Consequently, the setup of (\ref{result}) remains unchanged or does not change violently since neither the matter nor the gravitational sector can contribute
to $\Lambda_0$, and its determination from astrophysical observations involves thus no fine adjustments  of
distinct energy sources. In fact, as argued in \cite{banks1}, $\Lambda_0$ is not a calculable effective
parameter of the theory, instead it is the multiplicative inverse of the number of degrees of freedom in the fundamental theory.

A highly important property of the de Sitter spacetime is that energy is not positive definite everywhere; even if it is positive
in some portion of the spacetime it is negative in some other portion. This very feature guarantees that de Sitter space does not
support supersymmetry \cite{banks1}; any dynamical theory on it is necessarily nonsupersymmetric \cite{witten1}. Hence, the vacuum energy density induced by
the matter sector $\texttt{E}$ is around $\overline{M}_{Pl}^4$. This corresponds to the worst case, that is, the case with highest fine-tuning
in General Relativity \cite{ccp1,ccp2,nobbenhuis1}. Startlingly, however, this very scale for $\texttt{E}$, in view of (\ref{newton}), gives rise to the most natural
scheme for the mechanism. More explicitly, both the vacuum energy $\texttt{E}$ and $\texttt{L}^2 \simeq 1/\overline{M}_{Pl}^2$ are determined by
the Planck scale, as was already studied in (\ref{lkare}).

These discussions complete the proof of {\bf Proposition 6}. $\blacksquare$

\begin{figure}
\includegraphics[width=\textwidth, height=0.43\textheight]{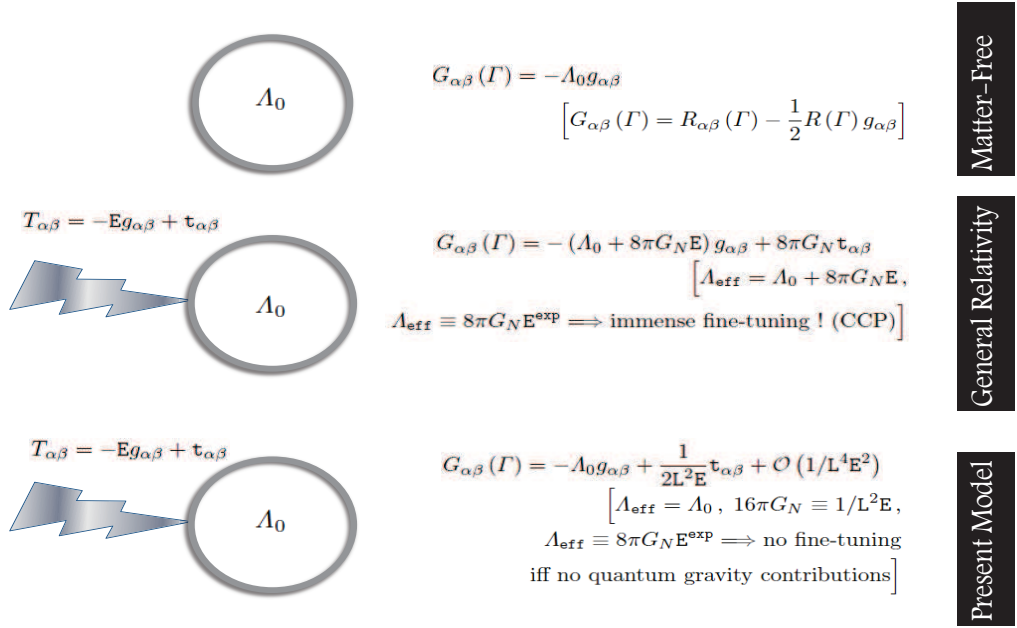}
\caption{\label{sekil1} A schematic summary of the mechanism (bottom) in comparison
with General Relativity (middle) with respect to the matter-free configuration (top) having
the cosmological constant $\Lambda_0$.}
\end{figure}

Having stated and proven the {\bf Propositions 1}--{\bf 6}, it could be useful to give a compact
overview of the mechanism. This is done in Fig. \ref{sekil1}. It provides a
comparative schematic summary of the mechanism. Shown at the top of the
figure is matter- and radiation-free spacetime with intrinsic curvature
$\Lambda_0$. The emanating gravitational field is described by (\ref{field0}).
Depicted in the middle of the figure is the Riemannian framework where switching on of
the energy-momentum tensor $T_{\alpha \beta}$ of matter and radiation gives rise to gravitational
field equations (\ref{field1}). This is the description in General Relativity, which suffers
from the CCP. The bottom of the figure describes the non-Riemannian framework proposed in the
present work. The response of the matter-free configuration to the inflow of the energy-momentum tensor
of matter and radiation is such that ({\it i}) the cosmological constant remains as in the
matter-free configuration, and thus, confrontation with observational results involves
no fine-tuning, and ({\it ii}) matter and radiation gravitate as in General Relativity
up to Planck-suppressed nonlocal terms.

\section{Yet More on the Mechanism}

Having formulated the mechanism, for completeness, it might be useful to discuss its a few important aspects, as we do below:
\begin{enumerate}

\item {\bf Propositions 1} and {\bf 5} give a detailed account of wherefrom the fundamental dynamical
equation (\ref{eqn2}) arises. It describes the dynamics of $\mathbb{T}_{\alpha\beta}$. By similarity to the matter-free gravitational
field equation (\ref{field0}), this dynamical equation can well be regarded as `matter-free gravitational field equation' in a
hyper-manifold with metric tensor $\mathbb{T}_{\alpha\beta}$ and connection coefficients $\between^{\lambda}_{\alpha\beta}$. The
connection $\between$ is nothing but the Levi-Civita connection on this manifold. In other words, $\between^{\lambda}_{\alpha\beta}$ is compatible
with $\mathbb{T}_{\alpha\beta}$, as clearly indicated by (\ref{eqn3}). Essentially, what the proposed mechanism does is to map the spacetime manifold
into this hyper-manifold
\begin{eqnarray}
{\cal{M}}\left(g,\Gamma\right)  \rightarrowtail {\cal{M}}\left(\mathbb{T},\between\right)
\end{eqnarray}
such that the two metrics are related through the energy-momentum tensor of matter on the spacetime manifold: $\mathbb{T}_{\alpha\beta}=\mathbb{T}_{\alpha\beta}\left[T\right]$.
It is this relation which generates the gravitational field equations (\ref{result}) through (\ref{relate}) in way free of the CCP, when gravity is
classical. The geometries of the manifolds are completely described by their metric tensors. The geometry of the hyper-manifold is defined by the
energy-momentum distribution of matter and radiation on the spacetime manifold. The whole mechanism would take a firm dynamical basis if the
hyper-manifold ${\cal{M}}\left(\mathbb{T},\between\right)$ and the underlying functional mapping $\mathbb{T}_{\alpha\beta}\left[T\right]$ can be constructed explicitly in a more fundamental theory {\it i.e.} string or M theories. Wishfully, it sounds appealing that the CCP becomes tameable in a geometry defined by the energetics of matter and radiation.

\item {\bf Proposition 6} argues that the de Sitter spacetime is essentially classical, and
thus, there cannot be any violent quantum gravitational corrections to (\ref{result}). However, this claim can
break down if the arguments backing it are falsified. A question that readily comes to mind is this: Can
the falsification originate from some properties of the mechanism proposed ? The answer is negative, to a high possibility. The reason is that the
mechanism is based on the dynamical equation (\ref{eqn2}) whose solution is obviously the de  Sitter
space. In other words, both manifolds ${\cal{M}}\left(\mathbb{T},\between\right)$  and
${\cal{M}}\left(g,\Gamma\right)$ are de Sitter spacetimes with their own metrics, and thus, it is
expected that the arguments pertaining to de Sitter space in {\bf Proposition 6} will continue to
hold for the proposed mechanism. Therefore, as for the General Relativity, the present mechanism
seems to involve essentially classical gravitation unless the arguments in {\bf Proposition 6} are falsified.

Let us suppose, for a moment, that either a way of quantizing the de Sitter gravity is found or de Sitter space
is shown to stem from string theory (as in, for instance, \cite{onemli3,onemli4}). Then, the arguments backing {\bf Proposition 6} break down. The
most important consequence of this will be the deposition of a new vacuum energy component by the quantum
fluctuations of graviton. This vacuum energy component arises either from the graviton tadpole in quantum de
Sitter gravity or from the quantum fluctuations in string or M-theoretic structures that compactify down to de Sitter space. In any case,
the vacuum energy density will be a quartically-divergent one indifferent from those generated by the matter loops, and
it regenerates the CCP with the same perplexity mentioned below (\ref{lameff}):
\begin{eqnarray}
\label{lam-grav}
\Lambda_0 \rightarrow \Lambda_0 + 8 \pi G_N \texttt{E}^{\texttt{grav}}
\end{eqnarray}
where $\texttt{E}^{\texttt{grav}}$ varies with the quartic power of the ultraviolet cutoff, and it becomes
unacceptably large compared to $\texttt{E}^{\texttt{exp}}$ for cutoffs above the neutrino mass scale. Therefore,
if de Sitter gravity is of quantum nature then the mechanism constructed in Sec. 2 is capable of taming
only the violent contributions of matter and radiation. In the presence of quantum fluctuations of
graviton, the impact of the present mechanism is limited to modifying the CCP to be a naturalness problem
`pertaining to the gravitational sector' alone; the contribution of the matter sector is utilized to generate
to the gravitational constant. In General Relativity, $\texttt{E}^{\texttt{grav}}$ gets embedded
into $\texttt{E}$ in the sense of (\ref{lameff}). Besides, attempts at
solving the CCP do not discuss `gravitational CCP' at all, simply because gravity is assumed
to be a classical phenomenon to start with \cite{ccp1,ccp2,nobbenhuis1,nobbenhuis2,review1,review3}.

In quantized gravity, the `gravitational CCP' in (\ref{lam-grav}) is to be tamed to achieve the naturalness.
To this end, one might consider a generalization of the mechanism in Sec. 2 where it can cover violent contributions from not only the
matter but also the gravitational sector. However, it is not clear if this can be done by a direct generalization. Another
possibility would be to embed an appropriate mechanism into the quantum theory of gravity itself, that is, the string theory
so that graviton loops or graviton-matter loops do not generate unacceptably large contributions \cite{review4}. One
here notes that the de Sitter space nonperturbatively arising from quantum fluctuations at the Planck scale possess
small quantum fluctuations \cite{onemli3,onemli4}, and violent contributions as in (\ref{lam-grav}) may not exist at all.

\item {\bf Proposition 4} brings up a novel length scale, $\texttt{L}$, not found in General Relativity. This parameter defines the gravitational constant $G_N$ via (\ref{newton}). It is inversely proportional to the total vacuum energy $\texttt{E}$ induced by the matter sector. The mechanism proposed treats it just as a parameter with no telling of its origin and possible connection with energetics and dynamics. According to {\bf Proposition 6}, it turns out to be of Planckian size in the de Sitter space induced by $\Lambda_0$. In one viewpoint, one can envision $\texttt{L}^2$ as a cross sectional area through which the covariantly-constant part of the energy-momentum tensor on ${\cal{M}}\left(g,\Gamma\right)$ is reflected to the metric tensor on $ {\cal{M}}\left(\mathbb{T},\between\right)$. This implies that $\texttt{L}^2$ might be associated with compactified dimensions in a higher dimensional formulation though it is not clear if such an approach, if any, can bring any new insight into the problem. In another viewpoint, one can consider it as a varying distance parameter that maintains the strength of gravitation against variations of the vacuum energy density  $\texttt{E}$. Indeed, in course of the evolution of the Universe, $\texttt{E}$ should have changed due to a series of phase transitions, the last one being the parton-hadron transition. The parameter $\texttt{L}$ thus works as a compensator for the changes in $\texttt{E}$ so that variations in the  gravitational constant remains within the observational limits in cosmology. Though the role of $\texttt{L}$ is rather clear in regard to (\ref{newton}), its origin is left unexplained by the mechanism. Needless to say, a proper understanding of $\texttt{L}$ will be possible if a more fundamental, possibly string theoretic,  formulation of the mechanism is accomplished.

\item {\bf Proposition 4} establishes gravitational field dynamics in the limit of small $\Theta/\Lambda$. This is an
excellent approximation for physical phenomena at ordinary energies. Indeed, even for ultra-high-energy cosmic rays with energies
in excess of $10^{19}\ {\rm eV}$, contribution of the remainder in (\ref{result}) is around $10^{-10}$ of the leading $\texttt{t}_{\alpha\beta}$
term.

For energetic systems, as energetic as to wander in the Planckian territory, the small $\Theta/\Lambda$ approximation breaks down,
and one is left with the exact equations (\ref{eqn2}). An immediate example of such ultra-high-energy systems is provided by cosmic inflation.
Inflation, exponential magnification of Planckian-size spacetime patches into regions some twenty orders of magnitude larger than the observable
part today, is the most celebrated mechanism for explaining the flatness, homogeneity and isotropy of the Universe.
The requisite outward pressure is provided by the inflaton field $\phi$ (whose energy density determines the expansion rate of the universe) which rolls
extremely slowly towards its minimum (due to the friction  induced by the expansion rate of the universe). For successful inflation,
energy density of the inflaton must start at Planckian values and diminish slowly as inflaton rolls down to its minimum. The fact that inflaton
acquires Planckian values is already a serious naturalness problem in General Relativity: the nonrenormalizable interactions are as important as the
renormalizable ones, and it is difficult to develop a consistent field-theoretic picture for inflaton dynamics \cite{inf1,inf2}.

The inflationary epoch offers a novel arena for studying the proposed mechanism in a regime in which the energy-momentum tensor of matter is of Planckian size.
In the present approach, a constant inflaton potential cannot inflate the Universe; it is used up for generating the gravitational constant as discussed in {\bf Propositions 4} and {\bf 6}. For an analysis of the inflation, instead of considering the exact equations (\ref{eqn2}), one may stick to (\ref{result}) by including higher powers in $\Theta/\Lambda$ to determine if the formalism respects the slow-roll regime of the inflationary epoch. To this end, for a slowly-rolling homogeneous inflaton field, one finds, after expanding $\phi(t)$ in Taylor series in $t$, that contributions of ${\cal{O}}\left(1/\Lambda^2\right)$ terms at the right-hand side of (\ref{result}) involve repeated derivatives $V\left(\phi\right)$ with respect to $\phi$. Therefore,
in (\ref{result}), the remainder cannot alter significantly the flatness of the potential. Though having  $\phi\sim \overline{M}_{Pl}$ continues to be a naturalness problem, the nonlocal contributions in view of the linear relation (\ref{relate}) can hardly modify the inflationary nature of the potential. One further notes that, letting $\phi$ develop a nonminimal coupling to $R\left(\Gamma\right)$ in $I_{matter}\left[g, \psi\right]$ in (\ref{basic-action}), where the resulting equation of motion  for $\phi$ will be identical to that in General Relativity, may improve the naturalness and other features\cite{inf3,inf4}.

This brief analysis, which obviously needs furthering for a precise determination of the model-dependent effects on the inflationary epoch, can be extended to cover the super-Planckian regime, $V\left(\phi\right) \simgt \overline{M}_{Pl}^4$, wherein the spacetime foam is expected to form. Though it requires quantum gravity to  understand such turbulent quantal spacetime structures, one may still extract information about the behavior of the mechanism proposed beyond the linear approximation (\ref{relate}) via a detailed analysis of (\ref{eqn2}).

\end{enumerate}
The points highlighted here should be taken as some representative topics which need further exploration. There is actually a whole list of phenomena that must be
dealt with within the present formalism: black holes, spacetime foam, grand unification, stringy scenarios, astrophysical phenomena and many more.
Essentially, having set up a framework where the CCP can have a resolution consistently gives rise to novel effects in other phenomena where
gravitational interactions are important.

\section{Conclusion}
Since its first solidification in \cite{ccp1}, the CCP has caused a huge literature
avalanche. The problem is a deeply perplexing one, and a resolution seems unlikely
to be found in the framework of quantum field theory and General Relativity. The literature
is widespread in terms of topics, scopes and methods of the attempts at understanding
the CCP. To mention a few of the main strategies, one notes various works utilizing
symmetry-based arguments \cite{nobbenhuis1,nobbenhuis2,review4,5,6,8,11,12,13,15,16,26,34,38,42,49},
relaxation mechanisms \cite{nobbenhuis1,review1,review2,review3,review4,review5,7,9,10,17,22,23,28,29,30,33,46,47,48,48p},
and modified gravitational theories \cite{nobbenhuis1,review1,review5,1,2,4,14,18,19,25,31,32,36,37,39,40,41,43,44,45,nima1,nima2}.
Given the well-understood part of Nature down to the terascale, a fundamental measure of the validity of any
proposition is its capability to degravitate the vacuum energy density (${\cal{O}}\left({M}_{W}^4\right)$ or higher)
induced by the matter sector. Therefore, each attempt at solving the CCP can be judged upon this minimal
requirement plus its validity and generality. Quite expectedly, any change designed for solving the
CCP must give rise to novel effects in quantum field theoretic and gravitational contexts, and it is
via these effects that one can hint in the true mechanism behind the tiny cosmological constant measured
\cite{astro1,astro2,astro3,astro4,astro5,astro6,astro7,astro8}.

Compared to the existing literature, the present work brings up a
novel, geometro-dynamical mechanism for a resolution  of the CCP.
The mechanism, based on the six propositions stated and proven in Sec. 2, enables
one to distinguish between Einstein's cosmological constant $\Lambda_0$
and the vacuum energy deposited by the quantal matter and radiation.
Thus, the vacuum energy of matter sector, instead of gravitating, facilitates the generation
of the gravitational constant. If gravity is a classical phenomenon, as has been
claimed for the de Sitter space generated by $\Lambda_0$, then the proposed mechanism can be
regarded to have naturalized the CCP since determination of $\Lambda_0$, in
isolation, from cosmological observations involves no fine-tuning at all.
On the other hand, if gravity is quantized, then CCP gets revived due to
graviton loops, and the mechanism proposed here is simply halted to offer any solution.
In this particular case, all that the present method can do is to modify the
nature of the CCP in that it becomes a naturalness problem pertaining to
the gravitational sector, only. Quantum gravity becomes the main obstacle in
searching for a way to suppress violent corrections from quantum fluctuations. However,
if de Sitter gravity remains as a purely classical structure in accord with various
claims then the mechanism proposed in this work gains a decisive status for the CCP.

The fundamental equations hypothesized are free of the CCP to start with. Reproduction
of correct gravitational dynamics, in a way free of the CCP, has been accomplished
by constructing a new geometry whose metric tensor is a linear, causal and nonlocal functional
of the energy-momentum tensor of the matter and radiation. If the covariantly-constant
part of this metric is large enough to facilitate a power series expansion, at the lowest order,
gravitational field equations are obtained by an appropriate definition of the gravitational constant.
Nevertheless, the original linear relation must be modified order by order in perturbation theory to reach
Bianchi consistency. The subleading terms obtained this way contain nonlocal pieces whose effects
can be crucial for description of Nature at super-Planckian energies where the aforementioned
power series expansion necessarily breaks down. For physical phenomena at ordinary energies,
however, the resulting dynamical equations are indistinguishable from the ones in General Relativity to
an excellent approximation. In heart, the mechanism makes critical
use of the dual nature of the metric tensor: It defines both the spacetime
geometry and energy-momentum tensor of the vacuum.

\section{Acknowledgments}
This work was partially supported by the Alexander von
Humboldt-Stiftung Friedrich Wilhelm Bessel-Forschungspreise and
by the Turkish Academy of Sciences through the GEBIP grant. The
author thanks the hospitality of the DESY Theory Group wherein
this work was started. He is grateful to Florian Bauer,
Rahmi G{\"u}ven and Hans Kastrup  for discussions, suggestions, and
a critical reading of an earlier version of the manuscript. He thanks
Wilfried Buchm{\"u}ller for useful conversations and suggestions. He also
thanks to anonymous referees for their illuminating criticism
and suggestions.

\end{document}